# Forecasting crude oil market volatility: can the Regime Switching GARCH model beat the single-regime GARCH models?

YUE-JUN ZHANG [a,b,], TING YAO [a,b], LING-YUN HE [c, d**]

[a] Business School, Hunan University, Changsha 410082, China
[b] Center for Resource and Environmental Management, Hunan University, Changsha 410082, China
[c] Institute of Resource, Environment and Sustainable Development Research, JiNan University, Guangzhou 510632, China
[d] Center for Futures and Financial Derivatives, College of Economics and Management, China Agricultural University, Beijing 100083, China

[] Corresponding author. Tel.: 86-731-88822899. E-mail: zyjmis@126.com (Prof. Dr. Yue-Jun Zhang).
[**] Corresponding author. Tel.: 86-13522821703. E-mail: lyhe@amss.ac.cn (Prof. Dr. Ling-Yun He).

**(1)The proposed running head (abbreviated form of the title):**

**FORECASTING CRUDE OIL MARKET VOLATILITY**

**(2)The name and mailing address, telephone, fax and e-mail numbers of the author to whom proof's should be sent:**

**LING-YUN HE**

**Institute of Resource, Environment and Sustainable Development Research, JiNan University, Guangzhou 510632, China. Tel.: +86 13522821703. E-mail: lyhe@amss.ac.cn**

# Forecasting crude oil market volatility: can the Regime Switching GARCH model beat the single-regime GARCH models?

YUE-JUN ZHANG [a,b,*], TING YAO [a,b], LING-YUN HE [c, d**]

[a] Business School, Hunan University, Changsha 410082, China
[b] Center for Resource and Environmental Management, Hunan University, Changsha 410082, China
[c] Institute of Resource, Environment and Sustainable Development Research, JiNan University, Guangzhou 510632, China
[d] Center for Futures and Financial Derivatives, College of Economics and Management, China Agricultural University, Beijing 100083, China

**Abstract**

In order to obtain a reasonable and reliable forecast method for crude oil price volatility, this paper evaluates the forecast performance of single-regime GARCH models (including the standard linear GARCH model and the nonlinear GJR-GARCH and EGARCH models) and the two-regime Markov Regime Switching GARCH (MRS-GARCH) model for crude oil price volatility at different data frequencies and time horizons. The results indicate that, first, the two-regime MRS-GARCH model beats other three single-regime GARCH type models in in-sample data estimation under most evaluation criteria, although it appears inferior under a few of other evaluation criteria. Second, the two-regime MRS-GARCH model overall provides more accurate volatility forecast for daily data but this superiority dies way for weekly and monthly data. Third, among the three single-regime GARCH type models, the volatility forecast of the nonlinear GARCH models exhibit greater accuracy than the linear GARCH model for daily data at longer time horizons. Finally, the linear single-regime GARCH model overall performs better than other three nonlinear GARCH type models in Value-at-Risk (VaR) forecast.

---

[*] Corresponding author. Tel.: 86-731-88822899. E-mail: zyjmis@126.com (Prof. Dr. Yue-Jun Zhang).
[**] Corresponding author. Tel.: 86-13522821703. E-mail: lyhe@amss.ac.cn (Prof. Dr. Ling-Yun He).

# Forecasting crude oil price volatility: can the Regime Switching GARCH model beat the single-regime GARCH models?

YUE-JUN ZHANG [a,b*], TING YAO [a,b], LING-YUN HE [c, d**]

[a] Business School, Hunan University, Changsha 410082, China
[b] Center for Resource and Environmental Management, Hunan University, Changsha 410082, China
[c] Institute of Resource, Environment and Sustainable Development Research, JiNan University, Guangzhou 510632, China
[d] Center for Futures and Financial Derivatives, College of Economics and Management, China Agricultural University, Beijing 100083, China

## 1. Introduction

The vital role of crude oil price in macro-economy is conclusive. As a crucial upstream product in the supply chain, the sudden and huge volatility of crude oil often leads to shock in productive capacity and further brings economic fluctuations. Meanwhile, the oil-importing and oil-exporting countries suffer economic instability due to the change of purchasing power. Moreover, crude oil is a peculiar commodity with political and financial properties, and some non-fundamental factors (such as speculation, geopolitics and US dollar currency rate) also make contribution to crude oil price movement. Hence, modeling and forecasting crude oil market volatility is a vital and complex issue in both commodity and financial markets (Elder and Serletis, 2011; Kilian and Vigfusson, 2011; Güntner, 2014; Zhang and Wang, 2013; Fan et al., 2008; Zhang et al., 2015).

Crude oil price volatility forecasts are often based on time-varying high frequency data and the samples of high volatility often have clustering feature. Hence, the Generalized Autoregressive Conditional Heteroscedasticity (GARCH) model proposed by Bollerslev (1986) is widely used for forecasting crude oil market volatility due to its good performance in capturing the time-varying feature of high frequency data (Wang and Wu, 2012; Kang et al., 2009; Agnolucci, 2009; Hou and Suardi, 2012; Marzo and Zagaglia, 2010; Mohammadi and Su, 2010). However, the standard GARCH model is intrinsically symmetric, and the forecast results with the standard GARCH model may be biased when the skewed time series are concerned (Franses and Dijk, 1996). To address this problem, some nonlinear and asymmetric GARCH models are proposed for crude oil price volatility forecast, like GJR-GARCH model by Glosten et al. (1993) and EGARCH model by Nelson (1991).

It should be noted that the GARCH type models above basically focus on one regime of crude oil price changes, whereas some specialists point out that the structural breaks in the variance process of the single-regime GARCH models often lead to a high persistence of volatility, because those models often fit the in-sample and out-of-sample data with the same pattern and ignore potential structural changes (Lamoureux and Lastrapes, 1990; Timmermann 2000). To address this problem, Cai (1994) and Hamilton and Susmel (1994) introduce the regime switching process (Hamilton, 1988, 1989) into the GARCH model, in order to consider potential structural breaks. In particular, the Markov Regime Switching based GARCH (MRS-GARCH) model permits the regimes in the Markov chain to have different GARCH behaviors, i.e., different volatility structures, in order to extend the GARCH model to the dynamic forms and realize better estimating and forecasting performance (Klaassen, 2002; Haas et al., 2004; Marcucci, 2005; Zhang and Wang, 2015; Zhang and Zhang, 2015).

Nonetheless, in terms of crude oil price volatility forecast, there still remain some intriguing problems to explore for GARCH type models. For instance, although the Markov Switching models are effective in capturing potential state transitions and the nonlinearities in crude oil price volatility, it is not certain whether the MRS-GARCH model definitely outperforms the single-regime GARCH models regarding crude oil price volatility forecast for all samples. Meanwhile, since the accuracy of GARCH type model forecasts often appears sensitive to data frequency and time horizon used to measure volatility (Manera et al., 2007; Zhang et al., 2015), it is interesting to see whether and how the results change when crude oil price volatility is calculated at different data frequencies, i.e., daily, weekly and monthly data, and different time horizons. Besides, in the empirical literature on the forecast performance comparison for the linear and nonlinear GARCH type models, no unanimous consensus has been reached.

Under this circumstance, we conduct some further research on crude oil market volatility forecast and the main contribution is as follows. First, we systematically evaluate and compare the forecast performance of three single-regime GARCH type models (including the standard linear GARCH model and two nonlinear GARCH type models) and the two-regime MRS-GARCH model for crude oil price volatility, and we also compare the forecast performance between the linear and nonlinear single-regime GARCH models. Second, we detect the effects of different data frequencies (daily, weekly and monthly data) and different time horizons on crude oil price volatility forecast. Third, besides the traditional evaluation criteria for forecast performance, we also employ the Value-at-Risk (VaR) based loss functions in order to obtain a more comprehensive evaluation, since crude oil volatility is a crucial input for the VaR model.

The remainder of this paper is organized as follows. Section 2 describes the data used for the empirical analysis. Section 3 introduces the models. Section 3 presents the empirical results and Section 4 concludes this paper.

## 2. Data description

This paper is about forecasting the volatility associated with the price of US West Texas Intermediate (WTI) crude oil. In order to obtain robust results, we use different data frequencies, i.e., daily, weekly and monthly data. The daily spot price data of WTI range from January 2, 2001 to April 23, 2015 with a total of 3843 observations. The first 3514 observations from January 2, 2001 to December 31, 2013 are selected as in-sample data for estimating the GARCH type models for crude oil price volatility, while the remaining 391 observations from January 2, 2014 to April 23, 2015 are selected as out-of-sample data for forecasting the volatility. The weekly spot crude oil price data range from January 5, 2001 to April 17, 2015 with a total of 798 observations. The first 730 observations from January 5, 2001 to December 27, 2013 are selected as in-sample data for estimation, while the remaining 68 observations from January 3, 2014 to April 17, 2015 are selected as out-of-sample data for forecast. Besides, the monthly data range from January 2001 to March 2015 with a total of 171 observations. The first 156 observations from January 2001 to December 2013 are selected as in-sample data and the remaining 14 observations from January 2014 to March 2015 as out-of-sample data. All the data come from the US Energy Information Administration (EIA).

In this paper, we define crude oil price return at time $t$ as $r_t = 100*[\log(p_t) - \log(p_{t-1})]$, where $p_t$ denotes crude oil price at time $t$. Meanwhile, the actual volatility of WTI crude oil price is defined as the squared return, and the daily crude oil prices, returns and actual volatility are shown in Figure 1.

**[Insert Figure 1 here]**

## 3. Methodology

### (1) The single-regime GARCH models

According to Bollerslev (1986) and Sadorsky (1999), the standard linear GARCH model for the WTI crude oil price returns can be specified as follows:

$$r_t = \delta + \varepsilon_t; \quad \varepsilon_t = \eta_t \sqrt{h_t}; \quad h_t = \alpha_0 + \alpha_1 \varepsilon_{t-1}^2 + \beta h_{t-1} \quad (1)$$

where $\alpha_0$, $\alpha_1$ and $\beta$ must be positive in order to guarantee a positive conditional variance, and $\alpha_1 + \beta < 1$ denotes the persistence of shocks to volatility. Following Klaassen (2002) and Haas et al. (2004), we adopt student's $t$ distribution for $\varepsilon_t$.

In order to consider the asymmetric leverage effect of crude oil price volatility, Glosten et al. (1993) propose the nonlinear GJR-GARCH model, and its variance equation is defined as Eq. (2).

$$h_t = \alpha_0 + \alpha_1 \varepsilon_{t-1}^2 [1 - I_{\{\varepsilon_{t-1}>0\}}] + \xi \varepsilon_{t-1}^2 I_{\{\varepsilon_{t-1}>0\}} + \beta_1 h_{t-1} \quad (2)$$

where $I_{\{\varepsilon_{t-1}>0\}}$ is an indicator function, and if $\varepsilon_{t-1} > 0$, $I_{\{\varepsilon_{t-1}>0\}} = 1$; otherwise, $I_{\{\varepsilon_{t-1}>0\}} = 0$.

Besides, given the standard GARCH model in Eq. (1) assumes that the effect of positive and negative information is symmetric, which may not completely accord with the market situation, Nelson (1991) proposes the Exponential GARCH (EGARCH) model to consider the asymmetric feature of asset price volatility, and the logarithm of the conditional variance of crude oil price returns can be written as Eq. (3).

$$\log(h_t) = \alpha_0 + \alpha_1 \left| \frac{\varepsilon_{t-1}}{h_{t-1}} \right| + \xi \frac{\varepsilon_{t-1}}{h_{t-1}} + \beta_1 \log(h_{t-1}) \tag{3}$$

**(2) The Markov Regime Switching GARCH (MRS-GARCH) model**

The main difference of the standard single-regime GARCH model and the multiple-regime MRS-GARCH model is that the parameters of the MRS-GARCH model are allowed to switch between different regimes following the Markov process. Specifically, the regime variable may switch according to a Markov process and the switching probability from regime $i$ at time $t-1$ to regime $j$ at time $t$ is equal to $P(s_t = j \mid s_{t-1} = i) = p_{ij}$.

Meanwhile, according to Klaassen (2002) and Haas et al. (2004), in this paper, we assume that the innovation $\varepsilon_t$ of the MRS-GARCH follows student's $t$ distribution with the degree of freedom of $\nu$, and the conditional mean, conditional variance following the GARCH process as well as the expectation of squared innovations can be specified as Eqs. (4), (5) and (6), respectively.

$$r_t = \mu_t^{(i)} + \varepsilon_t = \delta^{(i)} + \varepsilon_t; \quad \varepsilon_t = \eta_t \sqrt{h_t} \tag{4}$$

$$h_t^{(i)} = \alpha_0^{(i)} + \alpha_1^{(i)} \varepsilon_{t-1}^2 + \beta_1^{(i)} E_{t-1}\{h_{t-1}^{(i)} \mid s_t\} \tag{5}$$

$$E_{t-1}\{h_{t-1}^{(i)} \mid s_t\} = p_{ii,t-1}[(\mu_{t-1}^{(i)})^2 + h_{t-1}^{(i)}] + p_{ji,t-1}[(\mu_{t-1}^{(j)})^2 + h_{t-1}^{(i)}] - [p_{ii,t-1}\mu_{t-1}^{(i)} + p_{ii,t-1}\mu_{t-1}^{(i)}]^2 \tag{6}$$

where $i,j=1,2$ denotes the two regimes for the MRS-GARCH model [ii], $p_{ji,t} = \Pr(s_t = j \mid s_{t+1} = i, \varsigma_{t-1}) = \frac{p_{ji} \Pr(s_t = j \mid \varsigma_{t-1})}{\Pr(s_{t+1} = i \mid \varsigma_{t-1})} = \frac{p_{ji} p_{j,t}}{p_{i,t+1}}$ and $\varsigma_{t-1}$ represents the information at time $t-1$.

Then, in order to forecast the $k$-step-ahead volatility of crude oil prices, we sum the actual volatility of WTI during $k$ days. According to Klaassen (2002), the $k$-step-ahead volatility forecast at time $t$ can be written as Eq. (7).

$$\hat{h}_{t,t+k} = \sum_{\tau=1}^{k} \hat{h}_{t,t+\tau} = \sum_{\tau=1}^{k}\sum_{i=1}^{2} \Pr(s_\tau = i|\varsigma_{t-1})\hat{h}_{t,t+\tau}^{(i)} \qquad (7)$$

where $\hat{h}_{t,t+\tau}^{(i)} = \alpha_0^{(i)} + (\alpha_1^{(i)} + \beta_1^{(i)})E_t\{h_{t,t+\tau-1}^{(i)}|s_{t+\tau}\}$ is the $\tau$-step-ahead volatility forecast of regime $i$ at time $t$.

**(3) The evaluation criteria for forecast performance**

According to Marcucci (2005) and Wei et al. (2010), we use the loss functions as the evaluation criteria for the forecast results, including $MSE$, $MAD$, $QLIKE$ and $R^2LOG$, where $MSE$ and $MAD$ are the mean squared error and mean absolute error, respectively, $QLIKE$ represents the loss implied by a Gaussian likelihood and is introduced by Bollerslev et al. (1994), and $R^2LOG$ is the logarithmic loss function proposed by Pagan and Schwert (1990) and can penalize volatility forecast asymmetry in high and low level volatility.

Meanwhile, we also consider the directional predictive performance of various GARCH type models by using the Success Ratio (SR) and Directional Accuracy (DA) proposed by Pesaran and Timmermann (1992). Specifically, the value of SR changes between 0 and 1, and the larger the SR value appears, the higher directional predictive accuracy we can get. And the null hypothesis for the Directional Accuracy test is that the predicted and actual crude oil price volatility changes are independent. If the null hypothesis is rejected, we can say that the predicted crude oil price volatility direction is significantly relative to that of the actual volatility.

Besides, given that the volatility is often the input for the VaR estimation, we also employ the VaR estimation results to evaluate the forecast performance of the GARCH type models for crude oil price volatility. Specifically, we conduct three Log-likelihood rate (LR) tests, i.e., the LR test for unconditional coverage (LRuc), the LR test for independence (LRind) and the LR test for conditional coverage (LRcc) whose specific forms can be found from Brooks and Persand (2003). In the LR test for unconditional coverage (LRuc), the null hypothesis is that the failure probability is $\alpha$ at the significance level of $\alpha$; in the LR test for independence (LRind), the null hypothesis is that the failure process is independently distributed at the significance level of $\alpha$; and the null hypothesis for the LR test of correct conditional coverage (LRcc) is that the failure probability is $\alpha$ and the failure process is independent distributed at the significance level of $\alpha$.

## 4. Empirical results and discussions

### (1) In-sample estimation results

According to the models introduced in Section 3, we estimate the three single-regime GARCH models, i.e., the standard linear GARCH model, the nonlinear GJR-GARCH and EGARCH models, and the two-regime MRS-GARCH model, respectively. The results are shown in Tables 1 and 2, respectively, from which we have several findings.

For one thing, both the single-regime GARCH models and the two-regime MRS-GARCH model fit the daily crude oil prices very well. Specifically, based on the daily data, the regression coefficients in the three single-regime GARCH models in Table 1 and the two-regime MRS-GARCH model in Table 2 are basically statistically significant. Meanwhile, the standard linear GARCH model satisfies the second and fourth moment conditions provided by Ling and McAleer (2002a, 2002b). For another, when the data frequency gets lower from the daily data to the weekly and monthly data, the significance of the regression coefficients becomes weaker to some extent, for any GARCH type models concerned in this paper.

**[Insert Table 1 here]**

**[Insert Table 2 here]**

Meanwhile, the switching probabilities of the two-regime MRS-GARCH model are proven time-varying, among which the daily switching probabilities are shown in Figure 2. We can find that the regime switching occurs frequently. Specifically, in 2008, the switching probabilities of crude oil price volatility appeared not only higher than 0.5 but also lower than 0.5, i.e., there occurred regime switching. Meanwhile, most of the probabilities in that year were larger than 0.5, i.e., the WTI crude oil price stayed in the high volatility regime in most of the time in 2008, which is in line with the historical crude oil price data and crude oil market events due to the outbreak of global financial crisis.

**[Insert Figure 2 here]**

Besides, we evaluate the in-sample estimation performance of the three single-regime GARCH models and the two-regime MRS-GARCH model using the Akaike information criterion (AIC) principle and the four loss functions introduced in Section 3. The results are shown in Table 3, from which several important findings can be obtained.

For one thing, the two-regime MRS-GARCH model can beat other models concerned under most of the evaluation criteria used in this paper, although it is outperformed by other models under a few evaluation criteria. In other words, no model can completely outperform all of the other models in in-sample data estimation across all evaluation criteria. This finding is a little different from Fong and See (2002). They find that the regime switching GARCH model outperforms the single-regime GARCH models regardless of any evaluation criteria, because it can better capture and describe potential structural changes.

For another, for both daily and weekly data, the two-regime MRS-GARCH model provides more accurate in-sample estimation results than the three single-regime GARCH models overall. However, as for the monthly in-sample estimation, the superiority of MRS-GARCH model does not exist anymore. Specifically, as shown in Table 3, there are three and four out of five evaluation criteria that the MRS-GARCH model outperforms the single-regime GARCH models for the daily and weekly data, respectively, but for the monthly data, there is only one criterion (i.e., AIC) that the MRS-GARCH model can beat the single-regime GARCH models. One candidate interpretation for the monthly in-sample estimation inferiority is that there are eleven parameters in the two-regime MRS-GARCH model but the number of monthly in-sample observations proves much smaller than that of the daily and weekly in-sample observations. Meanwhile, another reason which cannot be excluded is that crude oil price volatility information has not been fully captured by the two-regime MRS-GARCH model due to the much smoother curve of monthly observations compared with the daily and weekly observations.

**[Insert Table 3 here]**

**(2) Out-of-sample forecast results**

A model which can well fit the in-sample data does not necessarily imply the model can accurately forecast the volatility for the out-of-sample data. Moreover, as accurate crude oil price volatility forecasts can effectively contribute to portfolio allocation along with risk investment and measurement, investors may focus more on the forecast performance of models (Wei et al., 2010). Hence, we forecast the volatility using the three single-regime GARCH models and the two-regime MRS-GARCH model based on different data frequencies (i.e., daily, weekly and monthly data). Furthermore, we also evaluate their forecast performance using the loss functions as mentioned above. The forecast results are shown in Table 4, from which we have several findings.

First, all the GARCH type models concerned provide relatively higher directional predictive accuracy for daily and weekly volatility forecast. Specifically, the SR values of all GARCH type models for the daily and weekly data are relatively higher, with the minimum 0.69 and 0.72 and the average 0.80 and 0.76 for the daily and weekly data, respectively. Moreover, the DA statistic of all the GARCH type models concerned are statistically significant at the 5% significance level for the daily and weekly data, which implies that the predicted crude oil price volatility direction is significantly relative to the actual volatility direction.

Second, the two-regime MRS-GARCH model overall performs better than the three single-regime GARCH models as for the volatility forecast for the daily data. However, this superiority of the two-regime MRS-GARCH model dies away at weekly and monthly frequencies. This finding is similar to Fong and See (2002), who argue that for the daily data, the volatility forecast performance of the single-regime GARCH models is outperformed by the MRS-GARCH model at 1-day horizon. Given that the out-of-sample daily observations range from January 2, 2014 to April 23, 2015 in this paper, and in the November of 2014 during this period, OPEC announced that there was no cutting of oil output, which then pushed the benchmark crude oil price to a new low. It is reasonable to suspect that this event changed the pattern of crude oil market's dynamic and the MRS-GARCH model's forecasting might get benefits from these out-of-sample data. For this reason, to verify whether the superiority of the MRS-GARCH model is sourced from the huge shocks in crude oil market, we select a common sub-sample to test the robustness of the MRS-GARCH model, which covers the period from January 2, 2001 to June 30, 2005 and was relatively stable without unexpected market events (Manera et al., 2007). The volatility forecast results are shown in Table A in the Appendix, which indicate that the MRS-GARCH model also has better forecast performance based on the common sample period overall. One candidate explanation of the superiority of the MRS-GARCH model in daily crude oil volatility forecast is that it appears more sensitive to shocks by switching the current regime, and higher frequency data can help to describe the nonlinearities of crude oil price volatility so as to better capture the regime switching.

Finally, among the three single-regime GARCH models, the volatility forecast of the two nonlinear GARCH models (i.e., the GJR-GARCH and EGARCH models) exhibit greater accuracy in daily data at longer time horizons than the linear GARCH model, but their superiority will be affected by data frequency and time horizon. Specifically, for the daily data, the nonlinear GARCH type models overall outperform the linear GARCH model in crude oil price volatility forecast, especially at longer time horizons (such as 10-day or 22-day ahead). However, for the weekly data, the linear GARCH model may provide more accurate volatility forecast results than the two nonlinear GARCH models at shorter time horizons (such as 1-week and 2-week ahead). This finding is in line with Wei et al. (2010), who find the nonlinear GARCH models perform better than the linear GARCH models for daily crude oil data. The superiority of the nonlinear GARCH models in daily volatility forecast may mainly result from their better performance in capturing the asymmetric leverage effect in volatility.

**[Insert Table 4 here]**

In addition, the VaR forecast results based on the daily, weekly and monthly data at the 5% significance level are reported in Table 5, from which we have several findings.

**[Insert Table 5 here]**

For one thing, the linear single-regime GARCH model overall performs better than the other three nonlinear GARCH type models in the VaR forecast. There are three evaluating statistics as shown in Table 5, i.e., LRuc, LRind and LRcc, which are used for testing the correct unconditional coverage, independence and conditional coverage, respectively. We can find that at the 5% significance level, for the daily data, only the linear GARCH model can pass the three LR tests at 1-day horizon. For the weekly data, no model passes the three LR tests. However, for the monthly data, all the four GARCH type models pass the three LR tests at the 5% significance level. This finding is in line with Dacco and Satchell (1999), who find that many nonlinear techniques give good in-sample fit, but they are usually outperformed in practical out-of-sample forecasting by simpler models due to the unreasonable mean squared error metric and possible over-fitting.

For another, the VaR forecast performance experiences greater accuracy at shorter time horizons. In fact, only for the monthly data, can all the GARCH type models completely pass the three LR tests at the 1-month horizon. For the daily data, only at the 1-day horizon, the GARCH model passes all the three LR tests and the EGARCH and MRS-GARCH model pass the LR test for independence. However, for the weekly data, no model really passes all the three LR tests at different time horizons.

## 5. Conclusions and future work

In this paper, we estimate and forecast the volatility for WTI crude oil prices at different data frequencies using three single-regime GARCH models (including the linear GARCH model and the nonlinear GJR-GARCH and EGARCH models) and the two-regime MRS-GARCH model. Taking into account what has been discussed above, we may safely come to the following main conclusions.

(1) The two-regime MRS-GARCH model may beat the three single-regime GARCH models under most of the evaluation criteria in in-sample estimation, although it is outperformed by the single-regime GARCH models under a few of other evaluation criteria. Meanwhile, the two-regime MRS-GARCH model overall provides more accurate in-sample estimation results than the single-regime GARCH models for both daily and weekly data. However, its superiority does not exist anymore when the monthly in-sample data is concerned.

(2) All the GARCH type models used in this paper provide relatively higher directional predictive accuracy for the daily and weekly crude oil price volatility forecast. Moreover, the two-regime MRS-GARCH model overall performs better than the three single-regime GARCH models for the daily data, but its superiority disappears at the weekly and monthly frequencies.

(3) Among the three single-regime GARCH type models, the two nonlinear GARCH models exhibit greater accuracy for the daily data at longer time horizons (such as 10-day and 22-day ahead) in crude oil price volatility forecast.

(4) All the GARCH type models provide greater forecast accuracy at shorter time horizons (such as 1-day ahead), and the linear single-regime GARCH model overall performs better than the nonlinear GARCH type models in the VaR forecast.

Besides, there is still some interesting work to be explored in the future. For instance, we can compare the GARCH type models with the hybrid forecast models. Meanwhile, the crude oil price volatility forecast performance can be evaluated based on both statistical values and traders' behaviors.

## Acknowledgements

We gratefully acknowledge the financial support from the National Natural Science

---

[i] The estimating period and the forecasting period are asymmetric since it is widely accepted that the in-sample period should be long enough and the out-of-sample period should be shorter. Besides, according to Behmiri and Manera (2015), GARCH forecasts are sensitive to the presence of outliers, but the empirical results in this paper indicate that the forecast errors are acceptable, i.e., outliers do not significantly change the final results. Therefore, we do not discard the outliers when the time periods are selected.

[ii] The literature focusing on regime switching of oil price volatility often adopts the two-regime or three-regime switching model. We test the number of regimes by the useful tool proposed by Hansen (1992, 1996) and we estimate the samples with both two-regime and three-regime switching models. The results indicate that most of the parameters in the three-regime switching model are not statistically significant, but the significance of the parameters in the two-regime switching model appears nice. Hence, we select the two-regime switching model in this paper.

# Tables and Figures

Table 1: The estimation results of single-regime GARCH models

| Model | $\delta$ | $\alpha_0$ | $\alpha_1$ | $\beta_1$ | $\xi$ | $\nu$ |
|---|---|---|---|---|---|---|
| *Panel A: Results for daily data* | | | | | | |
| GARCH | 0.0932*** | 0.1051*** | 0.0636*** | 0.9154*** | _ | 7.1651*** |
| | (2.8840) | (4.9290) | (7.4500) | (94.3580) | | (10.0360) |
| GJR-GARCH | 0.0812** | 0.1011*** | 0.0843*** | 0.9199*** | 0.0328*** | 7.2264*** |
| | (2.4880) | (5.1830) | (7.4740) | (104.1620) | (3.1900) | (10.0990) |
| EGARCH | 0.0651** | -0.0651*** | 0.1048*** | -0.0425*** | 0.9900*** | 7.1394*** |
| | (2.0060) | (-6.3450) | (7.3600) | (-4.4880) | (325.4720) | (10.2050) |
| *Panel B: Results for weekly data* | | | | | | |
| GARCH | 0.3122** | 0.4314** | 0.0705*** | 0.9029*** | _ | 8.9803*** |
| | (2.3490) | (2.4490) | (3.5250) | (40.2110) | | (4.3870) |
| GJR-GARCH | 0.2884** | 0.4586*** | 0.0887*** | 0.9049*** | 0.0395 | 9.2029*** |
| | (2.1270) | (2.6090) | (3.5710) | (38.2560) | (1.2750) | (4.4310) |
| EGARCH | 0.2400* | -0.0404 | 0.1308*** | -0.0527** | 0.9764*** | 9.9318*** |
| | (1.7810) | (1.0500) | (3.3420) | (2.3340) | (72.1500) | (4.1730) |
| *Panel C: Results for monthly data* | | | | | | |
| GARCH | 1.1175* | 10.0000 | 0.1645** | 0.6800*** | _ | 14.6418 |
| | (1.8830) | (1.2310) | (2.1540) | (3.652) | | (0.7610) |
| GJR-GARCH | 0.0705* | 2.0000 | 0.1438** | 0.8293*** | 0.1600 | 10.4753 |
| | (1.7750) | (0.9740) | (2.4980) | (13.0540) | (1.3340) | (1.1440) |
| EGARCH | 1.2577** | 0.9914 | 0.3241* | -0.1823* | 0.6945*** | 342.2000 |
| | (1.9650) | (1.6760) | (1.7700) | (1.7150) | (4.5260) | (0.0240) |

Note: The t-values of statistics are reported in parentheses. ***, ** and * denote the significance at the 1%, 5% and 10% levels, respectively.

Table 2: The estimation results of the two-regime MRS-GARCH model

| Coefficient | Daily | Weekly | Monthly | Coefficient | Daily | Weekly | Monthly |
|---|---|---|---|---|---|---|---|
| $\delta^{(1)}$ | 0.0599 | 1.3096*** | -11.9622*** | $\beta_1^{(2)}$ | 0.8906*** | 0.8098*** | 0.6047*** |
|  | (1.3680) | (3.4830) | (13.6650) |  | (30.9250) | (10.0570) | (3.9540) |
| $\delta^{(2)}$ | 0.1653*** | -1.0668* | 2.1920*** | $p$ | 0.9996*** | 0.7811*** | 0.2110 |
|  | (2.8130) | (1.9580) | (4.1790) |  | (1855.8300) | (7.2540) | (1.0940) |
| $\alpha_0^{(1)}$ | 0.0448** | 0.0001 | 0.0001 | $q$ | 0.9996*** | 0.7246*** | 0.9338*** |
|  | (2.2700) | (0.0001) | (0.0001) |  | (1608.790) | (6.5180) | (29.8860) |
| $\alpha_0^{(2)}$ | 0.4011*** | 0.0001 | 10.0001** | $\nu^{(1)}$ | 10.0291*** | 5.0481*** | 4.3847 |
|  | (2.7910) | (0.0001) | (2.0530) |  | (5.068) | (3.6040) | (0.6630) |
| $\alpha_1^{(1)}$ | 0.0640*** | 0.0001 | 1.0000*** | $\nu^{(2)}$ | 5.6071*** | 342.2000 | 326.7468 |
|  | (5.1910) | (0.0001) | (77.3960) |  | (7.378) | (0.3800) | (0.0210) |
| $\alpha_1^{(2)}$ | 0.0474*** | 0.1796** | 0.0309 |  |  |  |  |
|  | (3.3010) | (2.174) | (0.4000) |  |  |  |  |
| $\beta_1^{(1)}$ | 0.9266*** | 0.9009*** | 0.0001 |  |  |  |  |
|  | (65.0500) | (19.9860) | (0.0001) |  |  |  |  |

Note: The t-values of statistics are reported in parentheses. ***, ** and * denote the significance at the 1%, 5% and 10% levels, respectively.

Table 3: In-sample goodness-of-fit statistics

| Model | AIC Value | AIC Rank | MSE Value | MSE Rank | MAD Value | MAD Rank | QLIKE Value | QLIKE Rank | $R^2 LOG$ Value | $R^2 LOG$ Rank |
|---|---|---|---|---|---|---|---|---|---|---|
| *Panel A: Results for daily data* | | | | | | | | | | |
| GARCH | 4.386 | 3 | 226.271 | 3 | 6.440 | 2 | 2.594 | 3 | 7.239 | 2 |
| GJR-GARCH | 4.383 | 2 | 227.093 | 2 | 6.441 | 3 | 2.591 | 2 | 7.271 | 3 |
| EGARCH | 4.593 | 4 | 275.898 | 4 | 5.598 | 1 | 4.553 | 4 | 5.368 | 1 |
| MRS-GARCH | 4.381 | 1 | 224.016 | 1 | 6.445 | 4 | 2.583 | 1 | 7.291 | 4 |
| *Panel B: Results for weekly data* | | | | | | | | | | |
| GARCH | 5.585 | 4 | 8.150 | 4 | 18.674 | 4 | 3.769 | 4 | 6.228 | 4 |
| GJR-GARCH | 5.584 | 3 | 8.085 | 3 | 18.555 | 3 | 3.766 | 2 | 6.219 | 3 |
| EGARCH | 5.581 | 2 | 7.871 | 2 | 18.160 | 2 | 3.759 | 1 | 6.173 | 2 |
| MRS-GARCH | 5.576 | 1 | 7.681 | 1 | 17.751 | 1 | 3.767 | 3 | 5.974 | 1 |
| *Panel C: Results for monthly data* | | | | | | | | | | |
| GARCH | 7.136 | 4 | 28.825 | 1 | 65.022 | 1 | 5.280 | 4 | 6.965 | 1 |
| GJR-GARCH | 7.111 | 3 | 32.645 | 2 | 71.699 | 3 | 5.227 | 3 | 7.674 | 2 |
| EGARCH | 6.996 | 2 | 31.098 | 3 | 70.631 | 2 | 5.096 | 1 | 8.160 | 4 |
| MRS-GARCH | 6.927 | 1 | 48.595 | 4 | 85.773 | 4 | 5.181 | 2 | 7.688 | 3 |

Note: Rank=1, 2, 3, 4 represents the results from the best to the worst. AIC denotes the Akaike information criterion. MSE, MAD, QLIKE and $R^2 LOG$ are the statistical loss functions.

Table 4: Out-of-sample volatility forecast evaluation results

| Model | MSE | | MAD | | QLIKE | | $R^2LOG$ | | SR | DA |
|---|---|---|---|---|---|---|---|---|---|---|
| | Value | Rank | Value | Rank | Value | Rank | Value | Rank | | |
| *Panel A1: 1-day ahead volatility forecast for daily data* | | | | | | | | | | |
| GARCH | 113.08 | 3 | 1.12 | 4 | 2.30 | 3 | 7.82 | 3 | 0.74 | 5.92** |
| GJR-GARCH | 97.62 | 1 | 1.11 | 3 | 2.05 | 1 | 7.91 | 4 | 0.75 | 6.77** |
| EGARCH | 132.87 | 4 | 0.99 | 1 | 3.06 | 4 | 6.41 | 1 | 0.75 | 6.13** |
| MRS-GARCH | 111.82 | 2 | 1.11 | 2 | 2.18 | 2 | 7.45 | 2 | 0.74 | 5.92** |
| *Panel A2: 5-day ahead volatility forecast for daily data* | | | | | | | | | | |
| GARCH | 868.26 | 3 | 1.49 | 3 | 3.87 | 3 | 1.08 | 2 | 0.84 | 11.43** |
| GJR-GARCH | 781.95 | 1 | 1.48 | 2 | 3.82 | 2 | 1.09 | 3 | 0.84 | 11.53** |
| EGARCH | 1541.80 | 4 | 2.12 | 4 | 5.99 | 4 | 2.09 | 4 | 0.69 | 2.39** |
| MRS-GARCH | 860.32 | 2 | 1.40 | 1 | 3.81 | 1 | 0.93 | 1 | 0.84 | 11.43** |
| *Panel A3: 10-day ahead volatility forecast for daily data* | | | | | | | | | | |
| GARCH | 2182.00 | 3 | 1.86 | 2 | 4.57 | 3 | 0.65 | 3 | 0.808 | 13.36** |
| GJR-GARCH | 1912.03 | 1 | 1.86 | 3 | 4.54 | 2 | 0.64 | 2 | 0.88 | 13.16** |
| EGACH | 4855.93 | 4 | 3.06 | 4 | 7.04 | 4 | 2.12 | 4 | 0.71 | 3.40** |
| MRS-GARCH | 2155.57 | 2 | 1.69 | 1 | 4.51 | 1 | 0.49 | 1 | 0.88 | 13.36** |
| *Panel A4: 22-day ahead volatility forecast for daily data* | | | | | | | | | | |
| GARCH | 7577.91 | 3 | 2.78 | 3 | 5.43 | 3 | 0.59 | 3 | 0.85 | 11.88** |
| GJR-GARCH | 6173.85 | 1 | 2.70 | 2 | 5.39 | 2 | 0.56 | 2 | 0.84 | 11.67** |
| EGARCH | 20035.68 | 4 | 4.84 | 4 | 8.27 | 4 | 2.39 | 4 | 0.72 | 5.10** |
| MRS-GARCH | 7571.48 | 2 | 2.41 | 1 | 5.36 | 1 | 0.40 | 1 | 0.85 | 11.88** |
| *Panel B1: 1-week ahead volatility forecast for weekly data* | | | | | | | | | | |
| GARCH | 122.2054 | 1 | 2.0293 | 1 | 2.4524 | 1 | 12.0645 | 1 | 0.75 | 2.9064** |
| GJR-GARCH | 190.2055 | 3 | 2.3405 | 4 | 2.5704 | 4 | 12.7326 | 4 | 0.76 | 3.3163** |
| EGARCH | 234.6919 | 4 | 2.339 | 3 | 2.5247 | 3 | 12.2959 | 2 | 0.76 | 3.3163** |
| MRS-GARCH | 136.9691 | 2 | 2.1175 | 2 | 2.5023 | 2 | 12.4269 | 3 | 0.75 | 2.9064** |
| *Panel B2: 2-week ahead volatility forecast for weekly data* | | | | | | | | | | |
| GARCH | 486.8464 | 1 | 2.7256 | 1 | 3.258 | 2 | 8.8269 | 2 | 0.75 | 2.7258** |
| GJR-GARCH | 717.0674 | 3 | 3.0301 | 4 | 3.3436 | 4 | 9.3703 | 4 | 0.76 | 3.1753** |
| EGARCH | 777.3958 | 4 | 2.885 | 3 | 3.2515 | 1 | 8.7129 | 1 | 0.76 | 3.1753** |
| MRS-GARCH | 508.7151 | 2 | 2.7797 | 2 | 3.2792 | 3 | 8.9346 | 3 | 0.75 | 2.7258** |
| *Panel B3: 3-week ahead volatility forecast for weekly data* | | | | | | | | | | |
| GARCH | 1082.121 | 2 | 3.2066 | 2 | 3.7447 | 2 | 7.8461 | 3 | 0.79 | 3.7648** |
| GJR-GARCH | 1556.388 | 4 | 3.5839 | 4 | 3.8125 | 4 | 8.3814 | 4 | 0.80 | 4.1788** |
| EGARCH | 1489.453 | 3 | 3.2496 | 3 | 3.6775 | 1 | 7.5282 | 1 | 0.80 | 4.1788** |
| MRS-GARCH | 1062.455 | 1 | 3.1943 | 1 | 3.7509 | 3 | 7.783 | 2 | 0.79 | 3.7648** |
| *Panel B4: 4-week ahead volatility forecast for weekly data* | | | | | | | | | | |
| GARCH | 2016.383 | 2 | 3.6793 | 3 | 4.1136 | 3 | 8.7222 | 3 | 0.72 | 2.4895** |
| GJR-GARCH | 2722.053 | 4 | 4.0493 | 4 | 4.1665 | 4 | 9.1888 | 4 | 0.73 | 2.8745** |
| EGARCH | 2347.016 | 3 | 3.5224 | 1 | 3.9957 | 1 | 8.0972 | 1 | 0.73 | 2.8745** |
| MRS-GARCH | 1860.846 | 1 | 3.5432 | 2 | 4.1017 | 2 | 8.4092 | 2 | 0.72 | 2.4895** |
| *Panel C1: 1-month ahead volatility forecast for monthly data* | | | | | | | | | | |
| GARCH | 30.569 | 4 | 4.2351 | 4 | 5.3342 | 4 | 8.9279 | 3 | 0.67 | -0.2609 |
| GJR-GARCH | 20.4067 | 2 | 3.3036 | 1 | 4.8566 | 1 | 7.9032 | 1 | 0.72 | 0.8575 |
| EGARCH | 20.2764 | 1 | 3.877 | 3 | 4.9431 | 2 | 10.252 | 4 | 0.67 | 0.5065 |
| MRS-GARCH | 23.9739 | 3 | 3.5341 | 2 | 5.0752 | 3 | 8.515 | 2 | 0.72 | 0.8575 |

Note: Rank=1, 2, 3, 4 represents the results from the best to the worst. MSE, MAD, QLIKE and $R^2LOG$ are the statistical loss functions. SR and DA are the directional evaluation criteria. ** denotes the significance at the 5% level.

Table 5: Out-of-sample risk management evaluation: 5% VaR

| Model | LRuc | LRind | LRcc |
|---|---|---|---|
| *Panel A1: 1-day ahead volatility forecast for daily data* | | | |
| GARCH | 2.141 | 0.764 | 2.905 |
| GJR-GARCH | 591.231** | 4.112** | 595.343** |
| EGARCH | 31.857** | 1.882 | 33.739** |
| MRS-GARCH | 5.592** | 0.400 | 5.992** |
| *Panel A2: 5-day ahead volatility forecast for daily data* | | | |
| GARCH | 0.404 | 46.136** | 46.540** |
| GJR-GARCH | 665.848** | 113.842** | 779.690** |
| EGARCH | 77.462** | 68.141** | 145.604** |
| MRS-GARCH | 7.221** | 15.117** | 22.338** |
| *Panel A3: 10-day ahead volatility forecast for daily data* | | | |
| GARCH | 0.013 | 38.929** | 38.942** |
| GJR-GARCH | 697.983** | 225.508** | 923.491** |
| EGARCH | 132.701** | 146.178** | 278.879** |
| MRS-GARCH | 14.077** | 25.148** | 39.225** |
| *Panel A4: 22-day ahead volatility forecast for daily data* | | | |
| GARCH | 10.875** | 151.981** | 162.856** |
| GJR-GARCH | 730.470** | 283.098** | 1013.838** |
| EGARCH | 214.294** | 254.756** | 496.050** |
| MRS-GARCH | 9.548** | 134.685** | 144.233** |
| *Panel B1: 1-week ahead volatility forecast for weekly data* | | | |
| GARCH | 3.114 | 5.914** | 9.028** |
| GJR-GARCH | 150.392** | 5.720** | 156.112** |
| EGARCH | 0.697 | 4.817** | 5.514 |
| MRS-GARCH | 0.708 | 4.880** | 5.588 |
| *Panel B2: 2-week ahead volatility forecast for weekly data* | | | |
| GARCH | 9.075** | 8.725** | 17.799** |
| GJR-GARCH | 150.392** | 20.152** | 170.544** |
| EGARCH | 4.826** | 4.181** | 9.007** |
| MRS-GARCH | 0.106 | 7.023** | 7.129** |
| *Panel B3: 3-week ahead volatility forecast for weekly data* | | | |
| GARCH | 9.075** | 20.749** | 29.823** |
| GJR-GARCH | 170.752** | 36.006** | 206.757** |
| EGARCH | 6.822** | 11.394** | 18.216** |
| MRS-GARCH | 0.697 | 11.170** | 11.867** |
| *Panel B4: 4-week ahead volatility forecast for weekly data* | | | |
| GARCH | 20.271** | 31.083** | 51.354** |
| GJR-GARCH | 157.055** | 40.978** | 198.033** |
| EGARCH | 17.174** | 21.367** | 38.541** |
| MRS-GARCH | 4.826** | 22.544** | 27.370** |
| *Panel C1: 1-month ahead volatility forecast for monthly data* | | | |
| GARCH | 1.536 | 1.837 | 3.374 |
| GJR-GARCH | 1.536 | 0.681 | 2.217 |
| EGARCH | 0.08 | 0.159 | 0.239 |
| MRS-GARCH | 1.536 | 1.837 | 3.374 |

Note: LRuc, LRind and LRcc are the risk management loss functions based on VaR. ** denotes the significance at the 5% level.

Table A: Out-of-sample volatility forecast evaluation results for common sub-sample

| Model | MSE | | MAD | | QLIKE | | R²LOG | | SR | DA |
|---|---|---|---|---|---|---|---|---|---|---|
| | Value | Rank | Value | Rank | Value | Rank | Value | Rank | | |
| *Panel A1: 1-day ahead volatility forecast for daily sub-sample data* | | | | | | | | | | |
| GARCH | 1.8229 | 3 | 1.1419 | 3 | 2.3531 | 4 | 9.7461 | 2 | 0.45 | -3.1048 |
| GJR-GARCH | 1.7698 | 2 | 1.1332 | 2 | 2.3165 | 1 | 9.7559 | 3 | 0.54 | -0.7679 |
| EGARCH | 1.9044 | 4 | 1.1819 | 4 | 2.3422 | 2 | 9.9439 | 4 | 0.46 | -2.21 |
| MRS-GARCH | 1.6410 | 1 | 1.0680 | 1 | 2.3474 | 3 | 9.2097 | 1 | 0.40 | -4.1463 |
| *Panel A2: 5-day ahead volatility forecast for daily sub-sample data* | | | | | | | | | | |
| GARCH | 2.2049 | 3 | 1.2352 | 3 | 3.9586 | 3 | 0.5451 | 3 | 0.46 | -2.3213 |
| GJR-GARCH | 2.4464 | 4 | 1.3058 | 4 | 3.9717 | 4 | 0.5946 | 4 | 0.46 | -2.6337 |
| EGARCH | 1.9164 | 2 | 1.1300 | 2 | 3.9388 | 2 | 0.4822 | 2 | 0.46 | -1.7395 |
| MRS-GARCH | 1.4994 | 1 | 0.9584 | 1 | 3.9267 | 1 | 0.3631 | 1 | 0.41 | -3.2635 |
| *Panel A3: 10-day ahead volatility forecast for daily sub-sample data* | | | | | | | | | | |
| GARCH | 3.1902 | 3 | 1.5557 | 3 | 4.6522 | 3 | 0.364 | 3 | 0.57 | 1.2631 |
| GJR-GARCH | 3.5523 | 4 | 1.6541 | 4 | 4.6632 | 4 | 0.3983 | 4 | 0.50 | -0.6871 |
| EGARCH | 1.6329 | 2 | 1.0214 | 2 | 4.5965 | 1 | 0.2044 | 2 | 0.45 | -1.6777 |
| MRS-GARCH | 1.5613 | 1 | 0.9548 | 1 | 4.6064 | 2 | 0.1788 | 1 | 0.50 | -0.6871 |
| *Panel A4: 22-day ahead volatility forecast for daily sub-sample data* | | | | | | | | | | |
| GARCH | 6.6219 | 3 | 2.2907 | 3 | 5.4569 | 3 | 0.3058 | 3 | 0.55 | 0.7312 |
| GJR-GARCH | 7.1107 | 4 | 2.3776 | 4 | 5.4638 | 4 | 0.3247 | 4 | 0.54 | 0.3782 |
| EGARCH | 2.0011 | 1 | 1.1068 | 1 | 5.3862 | 1 | 0.0965 | 1 | 0.52 | -0.6808 |
| MRS-GARCH | 3.4468 | 2 | 1.3864 | 2 | 5.4428 | 2 | 0.1721 | 2 | 0.50 | -0.9152 |

Note: The in-sample observations are from January 2, 2001 to June 30, 2005 and the out-of-sample observations are from July 1, 2005 to December 30, 2005. MSE, MAD, QLIKE and $R^2LOG$ are the statistical loss functions.

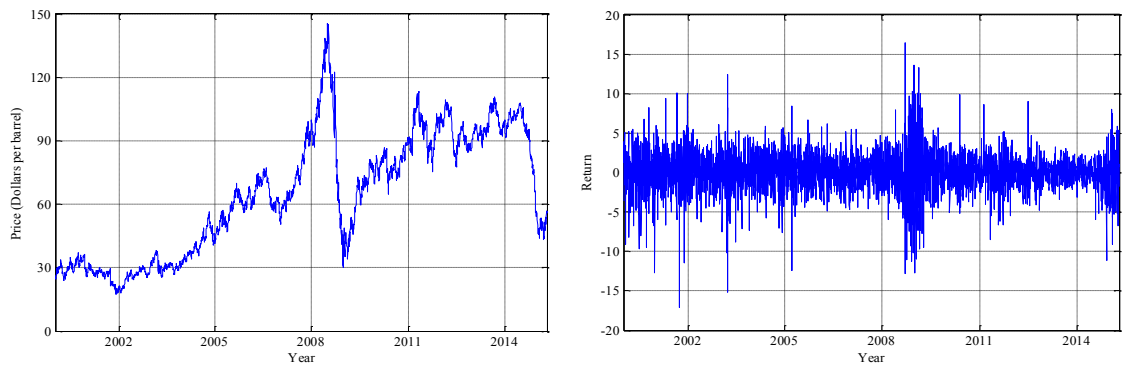

(a) Prices of WTI  (b) Returns of WTI

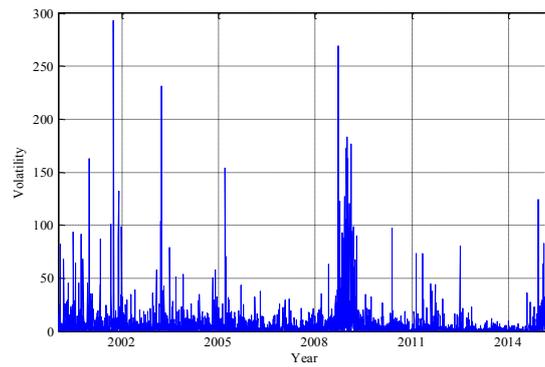

(c) Volatility of WTI

Figure 1　The daily prices, returns and volatility of WTI crude oil

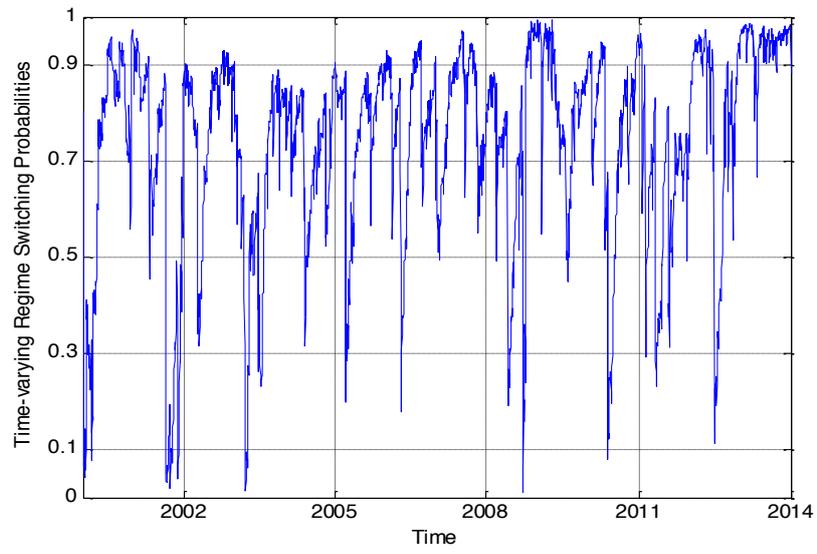

Figure 2    The time-varying regime switching probabilities for daily in-sample data